\documentclass[10pt,conference]{IEEEtran}

\usepackage[utf8]{inputenc}
\usepackage[T1]{fontenc}

\usepackage{cite}
\usepackage[cmex10]{amsmath}
\usepackage{amssymb}
\usepackage{amsthm}
\usepackage{graphicx}
\usepackage{booktabs}
\usepackage{hyperref}

\usepackage{algorithmic}
\usepackage{algorithm}
\usepackage[caption=false,font=footnotesize]{subfig}

 \usepackage{tikz}
\usetikzlibrary{positioning}
\usetikzlibrary{shapes, arrows.meta, decorations.pathmorphing, fit, backgrounds}

\begin{document}

\title{Topo-GS: Continuous Volumetric Embedding of High-Dimensional Data via Topological Gaussian Splatting}

\author{
    \IEEEauthorblockN{João Paulo Gois\IEEEauthorrefmark{1} and Luis Gustavo Nonato\IEEEauthorrefmark{2}}
    
    \IEEEauthorblockA{\IEEEauthorrefmark{1}Universidade Federal do ABC (UFABC), Santo André, Brazil\\
    Email: joao.gois@ufabc.edu.br}
    
    \IEEEauthorblockA{\IEEEauthorrefmark{2}Instituto de Ciências Matemáticas e de Computação (ICMC), \\ Universidade de São Paulo (USP), São Carlos, Brazil\\
    Email: gnonato@icmc.usp.br}
}

\maketitle

\begin{abstract}
Dimensionality reduction algorithms map high-dimensional data into visualizable 2D or 3D spaces, but traditionally rely on a discrete point-cloud paradigm. This discrete abstraction is susceptible to visual occlusion and artificial discontinuities, often failing to represent the continuous density of the underlying manifold. To address these limitations, we introduce Topo-GS, a framework that repurposes 3D Gaussian Splatting (3DGS) to cast multidimensional projection as a meshless volumetric reconstruction process. Instead of standard photometric losses, Topo-GS is driven by local geometric constraints. By solving orthogonal Procrustes targets, the optimization enforces an As-Rigid-As-Possible prior while explicitly aligning the spatial covariance of each Gaussian to the local tangent space. Recognizing that unrolling data of varying intrinsic dimensionalities requires distinct spatial treatments, we utilize a topology-aware strategy that tailors the loss formulation to preserve either continuous 1D trajectories or cohesive 2D surfaces. Quantitative and visual evaluations demonstrate that Topo-GS successfully transforms discrete scatter plots into continuous volumetric representations, where inherent projection distortions explicitly manifest as observable geometric variations, while preserving local topological fidelity comparable to discrete baselines.
\end{abstract}
 
\IEEEpeerreviewmaketitle

\section{Introduction}
Visualizing high-dimensional phenomena fundamentally relies on mapping complex data into interpretable low-dimensional spaces, typically 2D or 3D, that preserve the manifold structures~\cite{moon2019phate}. However, traditional dimensionality reduction relies primarily on discrete point clouds, an abstraction that can limit the capture of underlying manifold density and structural flow.

This discrete paradigm presents significant geometric and visual challenges. First, it is susceptible to structural fragmentation, such as topological tearing, in which continuous trajectories and manifolds are artificially shattered into disconnected clusters~\cite{moon2019phate}. Furthermore, as recently highlighted~\cite{jeon2025unveiling}, inherent distortions in dimensionality-reduction projections are often obscured in standard scatterplots, leading to visual representations that misrepresent the underlying high-dimensional data.

To address these limitations, we formulate dimensionality reduction as a volumetric reconstruction process. We present \emph{Topo-GS}, a framework that repurposes the 3D Gaussian Splatting (3DGS)~\cite{kerbl20233dgs} architecture as a geometric projection engine. Unlike standard 3DGS, which relies on a differentiable rasterizer and photometric loss to reconstruct scenes from 2D images, Topo-GS removes the rendering pipeline from the optimization loop. Instead, the Gaussian primitives are driven directly by local geometric constraints in the $n$-dimensional space. Using orthogonal Procrustes alignment~\cite{gower2004procrustes}, we enforce an As-Rigid-As-Possible prior~\cite{sorkine2007arap} to minimize local metric distortion, effectively translating projection stress into spatial deformations of ellipsoids.

The main contributions of this work are:
\begin{itemize}
    \item \textbf{Volumetric Manifold Reconstruction:} We formulate a dimensionality reduction technique based on 3DGS, transitioning the visual analytics paradigm from discrete points to a continuous, meshless volumetric representation.
    \item \textbf{Topology-Aware Optimization Framework:} We demonstrate that effective manifold unrolling requires dataset-specific regularizations. To address this, our framework provides a flexible loss formulation that incorporates robust error penalties and rigid constraints. By tuning covariance targets, the method can be tailored to preserve either 1D trajectories or intrinsic 2D surfaces, effectively mitigating structural collapse.
    \item \textbf{Geometric Manifestation of Distortion:} We translate inherent metric distortions into visually observable features by applying constraints that promote cohesive alignment. Consequently, projection stress manifests as variations in geometric thickness (for 2D surfaces) or structural buckling (for 1D trajectories).
\end{itemize}

\section{Related Work}
While linear techniques like PCA fail to capture nonlinear topologies, methods such as t-SNE~\cite{vandermaaten2008tsne} and UMAP~\cite{mcinnes2020umap} excel at preserving local neighborhoods. Advanced frameworks such as PHATE~\cite{moon2019phate} mitigate fragmentation by leveraging diffusion geometries, while recent global optimizations, such as the Enhanced Force-Scheme (EFS)~\cite{RosEFS}, improve layout convergence and mitigate local artifacts. However, because all these methods remain fundamentally tied to the discrete point-cloud paradigm, they inherently suffer from visual occlusion in dense regions and lack the capacity to represent the continuous volume of the underlying manifold.

To mitigate visual clutter and overdraw, existing methods abstract data into continuous density representations. Splatterplots~\cite{mayorga2013splatterplots} group dense discrete samples into bounded contours, while Continuous Scatterplots~\cite{bachthaler2008continuous, sharma2024continuous} derive density maps directly from spatially continuous input fields, bypassing discrete point sampling altogether.

Furthermore, to visualize projection distortion, previous work has overlaid 2D scatterplots with oriented glyphs~\cite{bian2020implicit} or structural uncertainty graphs~\cite{zhao2024mesoscopic} to highlight anisotropic stress. Although effective for highlighting distortions, these representations are fundamentally restricted to 2D encodings. Direct extensions to 3D environments are challenging: discrete 3D scatterplots inherently suffer from severe visual occlusion and depth ambiguity. On the other hand, representing the data as a continuous 3D volume typically relies on voxel grids, whose inherent discretization artifacts often obscure fine topological structures.

Projecting high-dimensional data robustly requires explicit geometric priors. As-Rigid-As-Possible (ARAP) modeling~\cite{sorkine2007arap} preserves local geometry by constraining transformations to orthogonal rotations, a concept successfully extended to abstract data projection by LAMP~\cite{joia2011lamp}, although still restricted to generating discrete point outputs.

To bridge the gap into a continuous, meshless domain, Topo-GS leverages 3D Gaussian Splatting (3DGS)~\cite{kerbl20233dgs}. While 3DGS is increasingly adopted for scientific volume rendering~\cite{tang2025ivrgs, ai2026nli4volvis} and extended to embed high-dimensional features for semantic querying~\cite{li2025langsplatv2}, these applications remain fundamentally anchored to physical scene coordinates. Instead of using 3DGS as a rendering tool optimized for photometric loss, Topo-GS repurposes it as a multidimensional projection engine. By integrating ARAP-inspired orthogonal targets directly into a topology-adaptive volumetric optimization, Topo-GS transitions the output from a point cloud to a continuous spatial field that explicitly encodes the mapping's geometric stress.

\section{Methodology}Topo-GS formulates dimensionality reduction as the joint optimization of $N$ anisotropic 3D Gaussians. Specifically, each high-dimensional sample $x_i \in \mathbb{R}^n$ is mapped to a volumetric primitive in $\mathbb{R}^3$. The projected spatial coordinate is defined by the Gaussian's mean ($\mu_i \in \mathbb{R}^3$), while the local topological deformation is encoded into its covariance ($\Sigma_i$) and rotation quaternion ($q_i$). Unlike photometric 3DGS, the opacity $\alpha_i$ is not optimized via gradient descent; instead, it is mapped to represent semantic features (e.g., energy states), enabling volumetric filtering. Secluding UMAP to a warm-start initialization to establish a spatial anchor, the Topo-GS engine relies on a multi-objective geometric loss function to iteratively refine the remaining parameters. Each term governs a distinct physical property of the manifold to translate high-dimensional affinities into a meshless volumetric field:\begin{itemize}\item \textbf{Local Rigidity ($\mathcal{L}_R$):} Drives the spatial positions ($\mu$), pulling Gaussian centers to unroll the manifold while preserving local metric distances via Procrustes alignment.\item \textbf{Covariance Alignment ($\mathcal{L}_C$):} Shapes the Gaussian geometry ($\Sigma$), forcing spherical kernels to flatten into ellipsoids that match the local tangent space.\item \textbf{Orientation Smoothing ($\mathcal{L}_O$):} Regularizes rotations ($q$), ensuring that neighboring ellipsoids align their orientations to form a smooth, continuous surface.\end{itemize}Recognizing that a single universal optimization regime is insufficient to accurately reconstruct datasets of varying intrinsic dimensionality, we implement a topology-adaptive optimization. This architecture allows the objective function to be explicitly tailored to the dataset's known structure, applying distinct physics to preserve either a 1D trajectory or an intrinsic 2D manifold. The total loss is a weighted sum of the three core components:\begin{equation}\mathcal{L}_{total} = \lambda{R} \mathcal{L}_{R} + \lambda{C} \mathcal{L}_{C} + \lambda{O} \mathcal{L}_{O},
\end{equation}
where empirically we set $\lambda_{R}=10$, $\lambda_{C}=10$, and $\lambda_{O}=2$.

\subsection{Local Affine Rigidity ($\mathcal{L}_{R}$)}
To preserve the intrinsic geometry during spatial unrolling, we enforce a local affine rigidity prior. While rooted in the As-Rigid-As-Possible (ARAP) surface modeling philosophy~\cite{sorkine2007arap}, our formulation is inspired by the Local Affine Multidimensional Projection (LAMP)~\cite{joia2011lamp}. Let $E_i \in \mathbb{R}^{k \times n}$ be the matrix of local displacement vectors for a $k$-NN neighborhood in the high-dimensional space $\mathbb{R}^n$, where each row is given by $x_j - x_i$, and $V_i(\mu) \in \mathbb{R}^{k \times 3}$ be the corresponding matrix of spatial edges in $\mathbb{R}^3$, where each row is given by $\mu_j - \mu_i$. To map the $n$-dimensional geometry to 3D, we solve the orthogonal Procrustes problem. By computing the Singular Value Decomposition of the cross-covariance matrix $V_i(\mu)^T E_i = U \Sigma W^T$, we discard the scaling matrix $\Sigma$ to enforce pure rigid rotation and extract the optimal orthogonal transformation $P_i = U W^T \in \mathbb{R}^{3 \times n}$. This yields the projection $V_{ideal} = E_i P_i^T$.

The application of the loss adapts to the underlying topological structure:
\begin{itemize}
\item \textbf{Intrinsic 2D Surfaces:} To enforce structural cohesion, we minimize the squared Frobenius norm of the edge transformations (MSE):
\begin{equation}\label{eqLrsur}
        \mathcal{L}_{R}^{(surface)} = \sum_{i=1}^{N} \| V_i(\mu) - V_{ideal} \|_F^2.
    \end{equation}
    This quadratic penalty minimizes deviations from the Procrustes target ($V_{ideal}$), enforcing an As-Rigid-As-Possible transformation. This local rigidity is essential for 2D topologies, ensuring the primitives tile cohesively to form a cohesive volumetric reconstruction and avoiding the introduction of artificial tearing.

\item \textbf{1D Trajectories:} In contrast, 1D sequences often contain isometric jumps compressed by the initial embedding. Applying an $L_2$ norm here would enforce rigidity and fracture the trajectory. Instead, we apply the Huber loss, which acts as a smoothed $L_1$ penalty. It penalizes large deviations linearly to absorb isometric stress, while transitioning to a quadratic minimum for stable convergence:
\begin{equation}\label{eqLrtraj}
        \mathcal{L}_{R}^{(trajectory)} = \sum_{i=1}^{N} \mathcal{H}_\beta \big( V_i(\mu) - V_{ideal} \big),
    \end{equation}
    where $\mathcal{H}_\beta$ denotes the Huber loss function with threshold~$\beta$.
\end{itemize}
 
\subsection{Adaptive Covariance Alignment ($\mathcal{L}_{C}$)}
Topo-GS forces the geometry of each Gaussian primitive, $\Sigma_i = R_i S_i S_i^T R_i^T$, to encapsulate the spatial distribution of its projected neighborhood. To ensure the resulting variance respects the intrinsic data affinities, we first compute normalized weights $w_{ij}$ for each neighbor $j \in \mathcal{N}(i)$ using a Gaussian kernel over the original high-dimensional distances~\cite{vandermaaten2008tsne, mcinnes2020umap} $d_{\mathbb{R}^n}$:
\begin{equation}
    w_{ij} = \frac{\exp \left( - d_{\mathbb{R}^n}(x_i, x_j)^2 / 2\sigma^2 \right)}{\sum_{k \in \mathcal{N}(i)} \exp \left( - d_{\mathbb{R}^n}(x_i, x_k)^2 / 2\sigma^2 \right)}.
\end{equation}

Using these high-dimensional weights, we construct a target 3D covariance matrix $C_{target} = \sum_{j} w_{ij} (v_{ij} v_{ij}^T)$. Geometrically, the outer product $v_{ij} v_{ij}^T$ captures the directional spread of a single neighbor. Weighting this sum ensures that points originally closer in $\mathbb{R}^n$ dictate the principal axes of the Gaussian in $\mathbb{R}^3$. The choice of the 3D edge vectors $v_{ij}$ determines the topological regime:

\begin{itemize}
    \item \textbf{Intrinsic 2D Surfaces:} The edge vectors are derived from the current 3D spatial positions ($v_{ij} = \mu_j - \mu_i$). Unlike the linear Procrustes target, which is strictly flat, the current 3D neighborhood captures the emergent extrinsic curvature of the manifold during optimization. Building $C_{target}$ from these spatial edges allows the Gaussian to adapt to this local tangent space, tiling cohesively to form a continuous volumetric shell.
    
    \item \textbf{1D Trajectories:} In contrast, a 1D sequence often buckles and compresses in 3D space to absorb isometric stress. If we constructed $C_{target}$ using these distorted spatial edges, the primitives would clump into isotropic blobs, reinforcing the error. Instead, we extract the edge vectors directly from the analytical target ($v_{ij} = v^{ideal}_{ij}$, where $v^{ideal}_{ij}$ corresponds to the respective row in $V_{ideal}$). This forces the Gaussian to ignore the local 3D folding and align its principal axis with the intrinsic 1D path, forming a continuous tube.\end{itemize}

Finally, to force the optimizable Gaussian geometry ($\Sigma_i$) to match this spatial footprint, the loss is computed as:
\begin{equation}\label{eqLc}
    \mathcal{L}_{C} = \frac{1}{N} \sum_{i=1}^{N} \| \Sigma_i - C_{target} \|_F^2.
\end{equation}

\subsection{Topological Orientation Smoothing ($\mathcal{L}_{O}$)}
To promote volumetric cohesion and mitigate misaligned intersections between neighboring ellipsoids, we introduce a rotation smoothing regularization. By normalizing the quaternion vectors $q_i$, we penalize the angular misalignment between neighboring Gaussian orientations, encouraging them to tile cohesively along the manifold:
\begin{equation}\label{eqLo}
    \mathcal{L}_{O} = \frac{1}{N} \sum_{i=1}^{N} \frac{1}{k} \sum_{j \in \mathcal{N}(i)} \left(1 - (q_i \cdot q_j)^2\right).
\end{equation}

\subsection{Scale Control and Manifold Reconstruction}
To maintain optimization stability, we apply regime-specific constraints on the Gaussian log-scales throughout training. 

\textbf{Training Phase:} For 2D surfaces, we impose a restrictive clamp ($\max = -0.5$) on the log-scales to ensure the primitives remain as thin planar elements, preventing volumetric overlap that could destabilize the spatial unrolling. In contrast, 1D trajectories utilize a relaxed boundary ($\max = 1.0$), allowing the Gaussians to elongate and bridge temporal gaps in the sequence.
    
\textbf{Visualization Phase:} Upon convergence, a decoupled adjustment ensures visual continuity. For the 2D regime, the converged log-scales are expanded by a constant factor (e.g., $+0.5$), followed by a uniform opacity saturation ($\alpha \approx 0.85$). In contrast, 1D trajectories utilize a non-linear power-law mapping to enhance the visual resolvability of dynamic states:
\begin{equation}\label{eqalpha}
    \alpha_i = \alpha_{min} + (\alpha_{max} - \alpha_{min}) \cdot \hat{E}_i^{1.5},
\end{equation}
where $\hat{E}_i \in [0, 1]$ is the normalized energy state and the range is set to $[\alpha_{min}=0.15, \alpha_{max}=0.95]$. This scaling maximizes visual contrast along the sequence, ensuring that the structural flow remains interpretable despite 3D projection stress.

This decoupled strategy ensures that the manifold structure learned during training is preserved, while the resulting render reconstructs a cohesive volume without visual fragmentation.

\subsection{Optimization Strategy}
To balance computational efficiency and convergence stability, the orthogonal Procrustes targets ($V_{ideal}$) are updated dynamically using a two-phase lazy strategy. The targets are updated every $\tau = 15$ steps to guide the global unrolling, but are completely frozen halfway through the optimization (at epoch $M_{stop} = 100$). This target freezing guarantees that during the second half of training (up to $M = 200$), the spatial anchors remain static, allowing the optimization to fine-tune exclusively the local Gaussian geometry ($\Sigma_i, q_i$) without moving the underlying topological ground. The complete optimization loop of the Topo-GS framework is summarized in Algorithm~\ref{alg:topogs}.

\begin{algorithm}
\caption{Topo-GS Optimization Engine}
\label{alg:topogs}
\begin{algorithmic}[1]
\REQUIRE High-dim data $\mathbf{X} \in \mathbb{R}^{N \times n}$, $k$-NN graph $\mathcal{G}$, epochs $M$, lazy interval $\tau$, freeze epoch $M_{stop}$
\ENSURE Optimized 3D Gaussians (means $\mu$, covariances $\Sigma$, quaternions $q$)
\STATE Initialize $\mu$ via UMAP warm-start
\STATE Initialize $q \leftarrow [1, 0, 0, 0]$, log-scales $\leftarrow s_{init}$
\FOR{$\text{step} = 1$ \TO $M$}
    \IF{$\text{step} \bmod \tau == 0$ \AND $\text{step} \le M_{stop}$}
        \STATE Update Procrustes orthogonal targets $V_{ideal}$ via SVD
    \ENDIF
    \STATE Compute 3D spatial edges $V_i(\mu)$
    \STATE Compute Local Rigidity Loss $\mathcal{L}_R$ (Eq.~\eqref{eqLrsur} or~\eqref{eqLrtraj})
    \STATE Compute target covariance $C_{target}$ from local tangent space
    \STATE Compute Covariance Alignment Loss $\mathcal{L}_C$ (Eq.~\eqref{eqLc})
    \STATE Compute Orientation Smoothing Loss $\mathcal{L}_O$ (Eq.~\eqref{eqLo})
    \STATE $\mathcal{L}_{total} \leftarrow \lambda_R \mathcal{L}_R + \lambda_C \mathcal{L}_C + \lambda_O \mathcal{L}_O$
    \STATE Backpropagate $\mathcal{L}_{total}$ and update $(\mu, \text{scales}, q)$
    \STATE Apply regime-specific clamp on log-scales
\ENDFOR
\STATE Apply visualization opacity mapping (Eq.~\eqref{eqalpha})
\end{algorithmic}
\end{algorithm}
 
\section{Results}
\begin{figure*}[htb]
    \centering
    \includegraphics[width=0.32\textwidth]{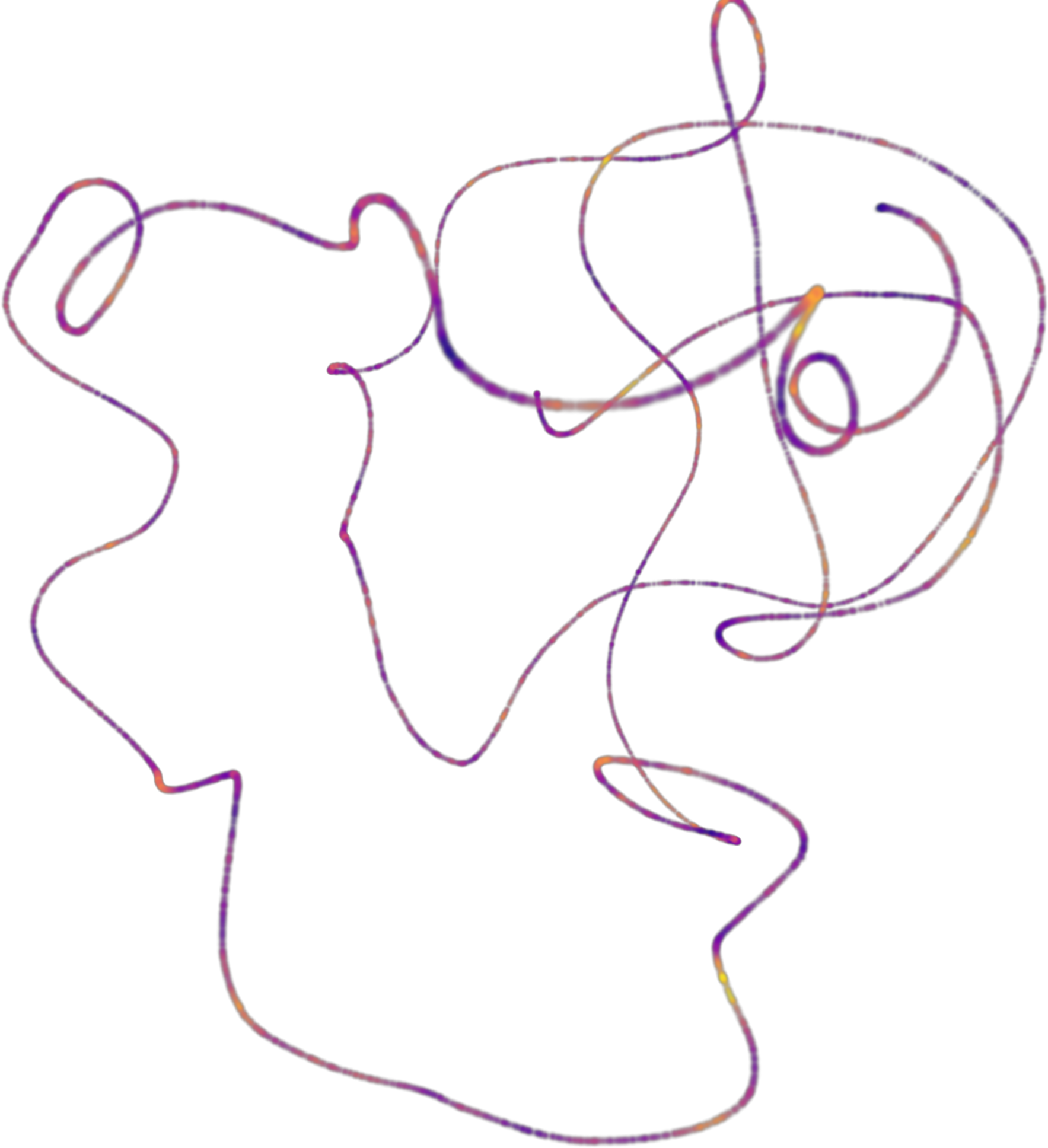}
    \includegraphics[width=0.32\textwidth]{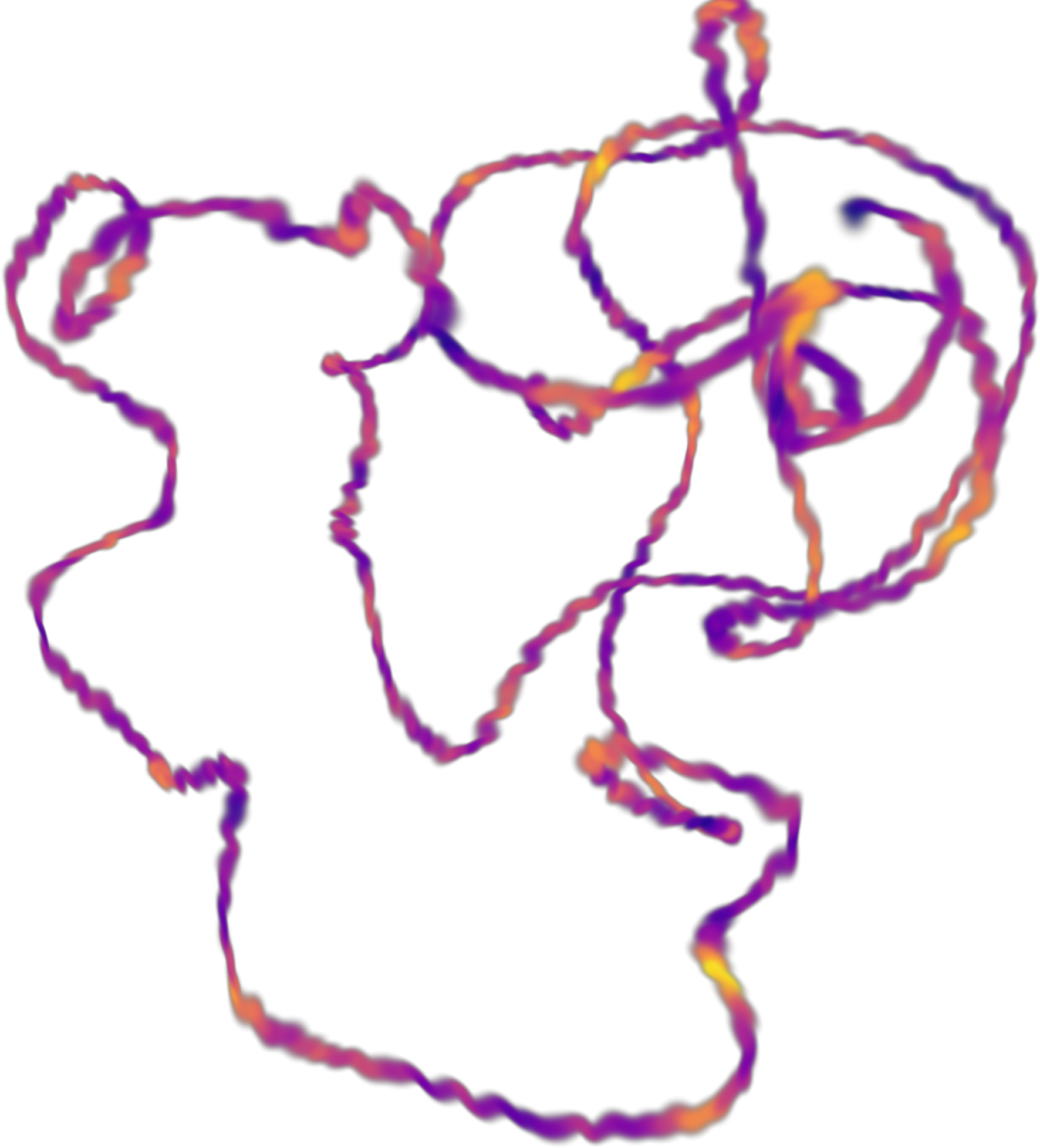}
    \includegraphics[width=0.32\textwidth]{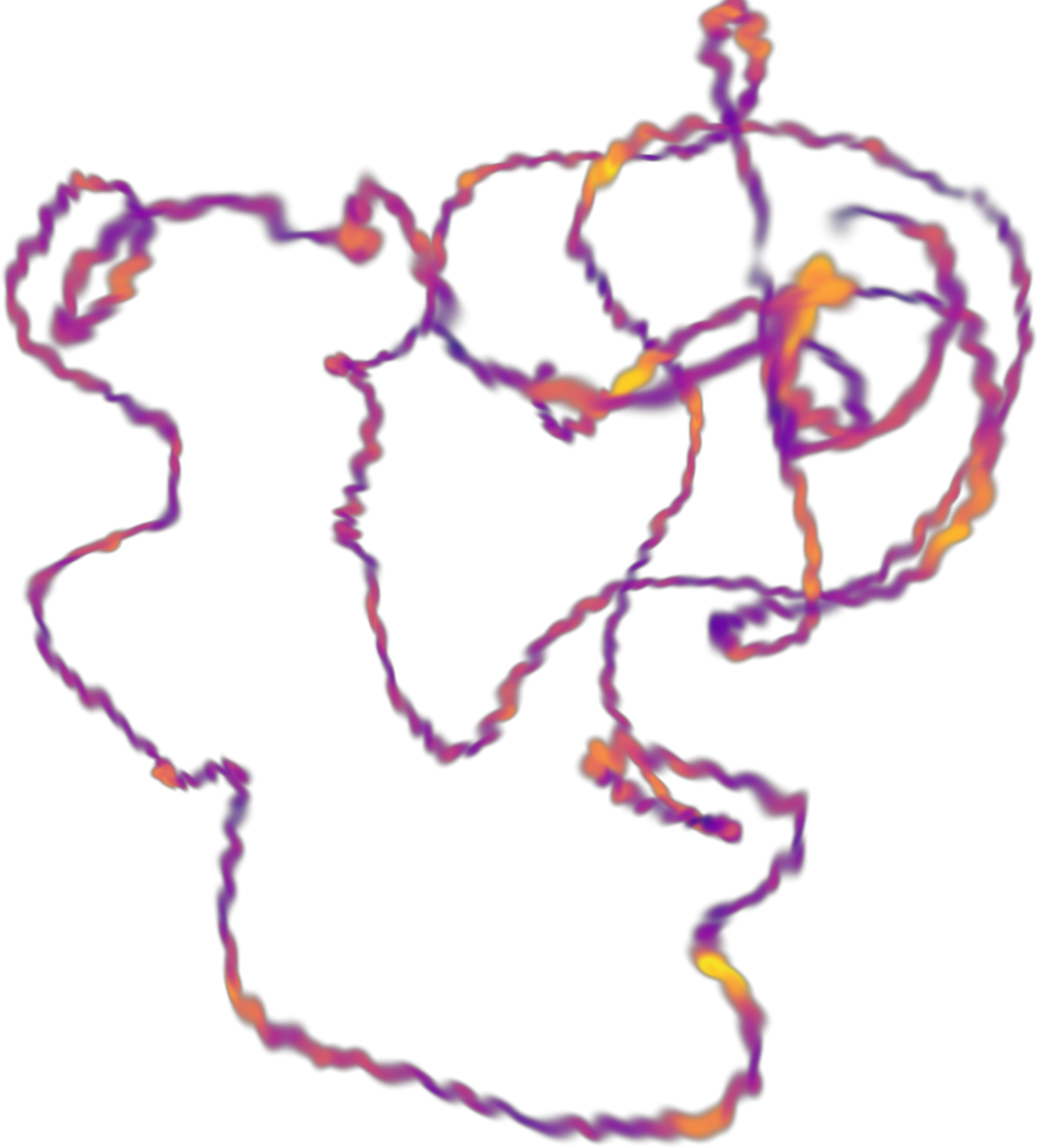}    
    \includegraphics[width=0.32\textwidth]{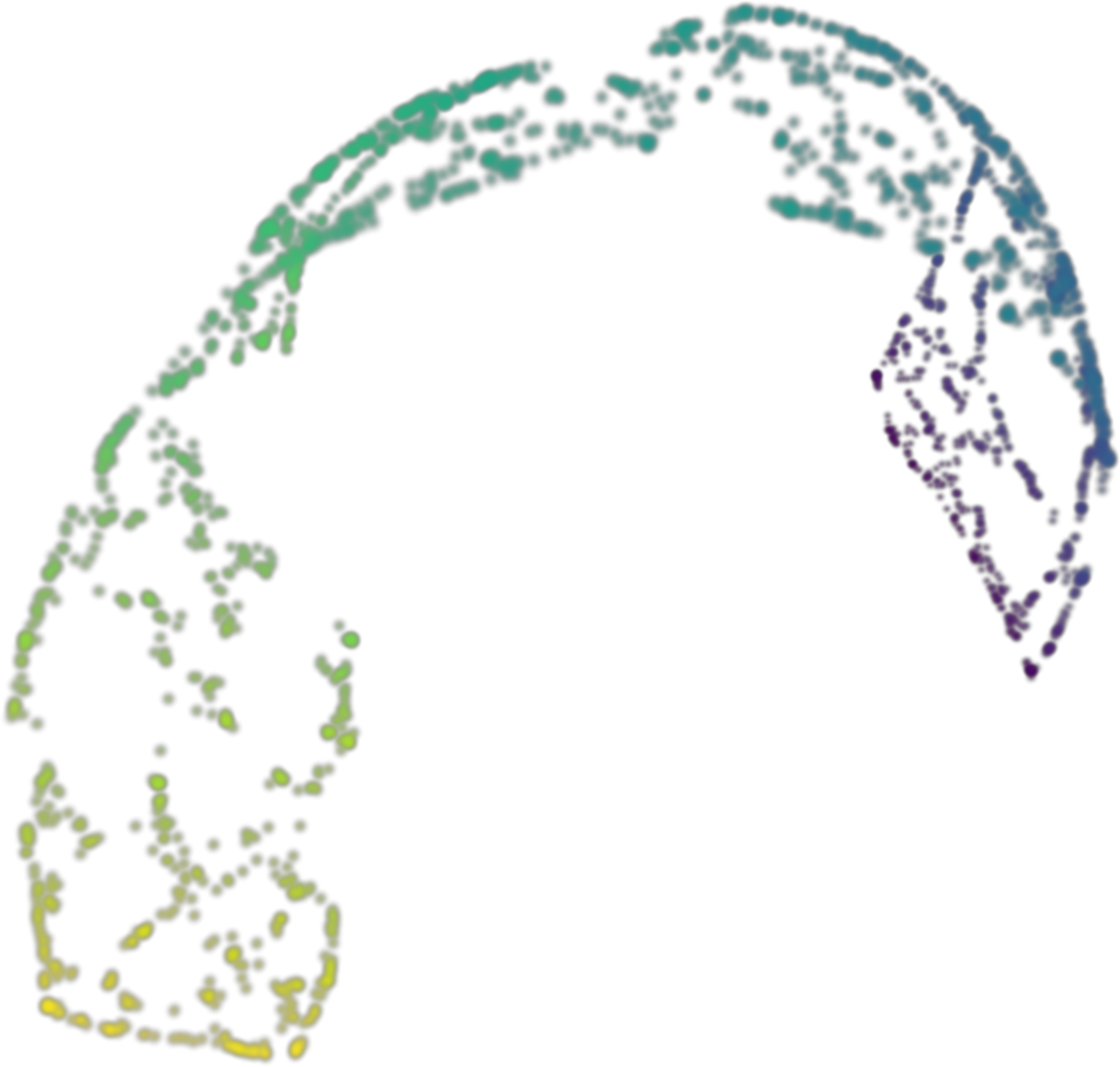}
    \includegraphics[width=0.32\textwidth]{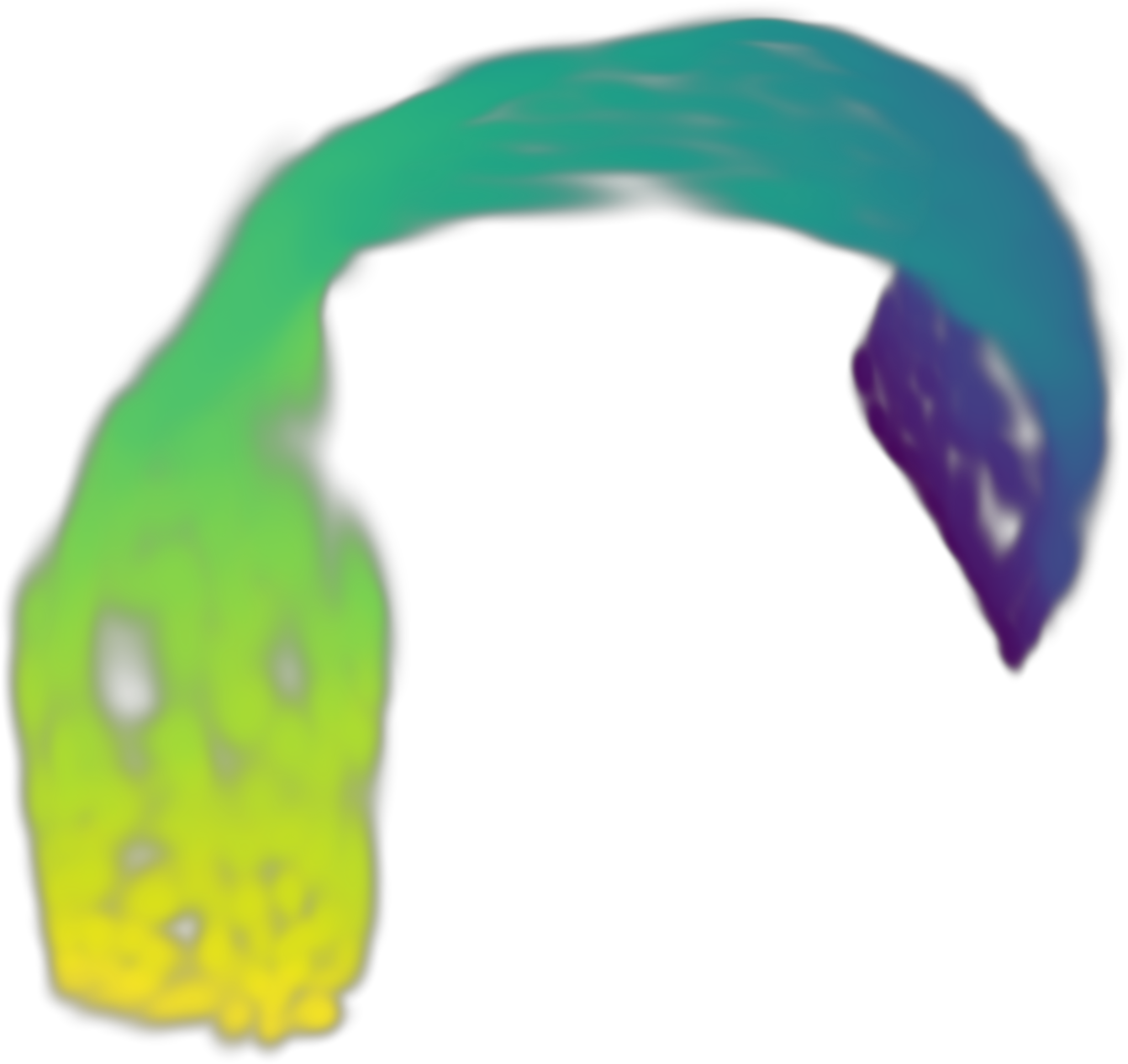}
    \includegraphics[width=0.32\textwidth]{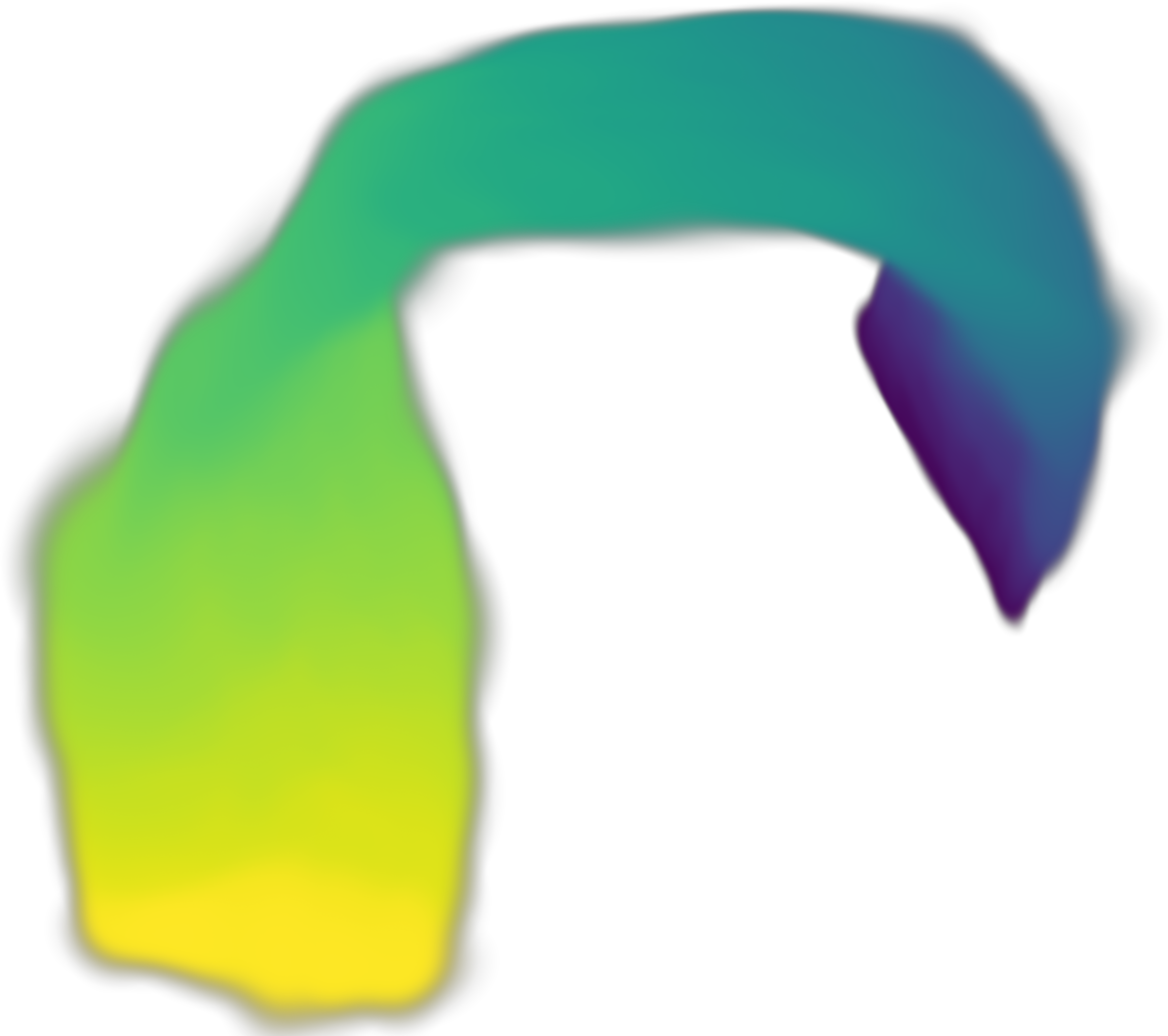} 
\caption{Visual evolution of the 1D Uracil trajectory (top) and 2D Swiss Roll (bottom). The initial state (left) shows the UMAP baseline. During optimization (center), Gaussian primitives adapt to the local geometry to bridge structural gaps, resulting in a cohesive manifold reconstruction (right).}
    \label{fig:evolution}
\end{figure*}

We evaluate Topo-GS on two benchmarks with distinct intrinsic dimensionalities: the MD17 Uracil dataset (1D trajectory) and the Swiss Roll (2D surface). Results demonstrate that the proposed volumetric reconstruction bridges structural gaps while preserving the metric fidelity of the original embedding.

\subsection{Qualitative Evaluation and Volumetric Rendering}
Figure \ref{fig:evolution} illustrates the geometric transition from discrete UMAP seeds (Epoch 0) to a cohesive manifold reconstruction.
 
Traditional dimensionality reduction methods often suffer from point collapse in high-density regions, where multiple high-dimensional samples are projected onto nearly identical 3D coordinates. Topo-GS mitigates this effect through structural decoupling: by assigning a learnable covariance $\Sigma_i$ to each primitive, the optimization encourages Gaussians to occupy distinct spatial footprints. This spatial partitioning enhances visual resolvability, preserving the local flow of the data that is typically obscured in discrete point-based embeddings.

\textbf{1D Trajectories (Uracil).} The MD17 Uracil dataset represents a 1D sequence of molecular dynamics. While traditional discrete embeddings often collapse this temporal sequence into thin, dimensionless lines (Fig. \ref{fig:evolution}, top left), Topo-GS explicitly visualizes the projection stress. Because embedding dynamic state-spaces into $\mathbb{R}^3$ forces severe isometric compression, the trajectory regime (via Huber loss) absorbs this stress by geometrically buckling the continuous tube (Fig. \ref{fig:evolution}, top right) to accommodate the volumetric mass of the ellipsoids. These spiraling structures are the geometric manifestation of high-dimensional variance, providing a direct visual cue of the embedding's local distortion.

For the 2D Swiss Roll dataset, standard projection algorithms produce a discrete point cloud, and lower sampling density results in visual discontinuities (Fig. \ref{fig:evolution}, bottom left). The Topo-GS surface regime utilizes 3D spatial covariance to align the primitives along the local tangent plane. The anisotropic Gaussians flatten and overlap, bridging these gaps to reconstruct the manifold as a cohesive volumetric shell (Fig. \ref{fig:evolution}, bottom right).

\subsection{Quantitative Evaluation}
To verify that the volumetric reconstruction preserves the underlying manifold structure, we evaluated the converged Gaussian centers ($\mu$) against the high-dimensional data using Kruskal's Stress-1, Trustworthiness, and Continuity ($k=15$)~\cite{VennaKaski2001}. As shown in Tab.~\ref{tab:quant_metrics}, Topo-GS maintains neighborhood preservation scores comparable to the UMAP baseline. The minimal variation in Kruskal's Stress-1 (an increase of less than $0.3\%$) suggests that the additional geometric constraints required for volumetric tiling do not significantly degrade global distance preservation. This marginal trade-off indicates that the model can prioritize structural cohesion and rotational alignment while maintaining the underlying embedding's metric integrity.

\begin{table}[htbp]
    \centering
\caption{Quantitative comparison of topological and metric preservation.}
    \label{tab:quant_metrics}
    \resizebox{\columnwidth}{!}{%
    \begin{tabular}{llccc}
        \toprule
        \textbf{Dataset} & \textbf{Method} & \textbf{Stress-1 ($\downarrow$)} & \textbf{Trustworthiness ($\uparrow$)} & \textbf{Continuity ($\uparrow$)} \\
        \midrule
        \textbf{Uracil (1D)} & UMAP 3D & \textbf{0.4555} & 1.0000 & 0.9994 \\
                             & Topo-GS & 0.4559 & \textbf{1.0000} & \textbf{0.9995} \\
        \midrule
        \textbf{Swiss Roll (2D)} & UMAP 3D & \textbf{0.4602} & 0.9992 & 0.9981 \\
                                 & Topo-GS & 0.4615 & \textbf{0.9999} & \textbf{0.9998} \\
        \bottomrule
    \end{tabular}%
    }
\end{table} 
The quantitative results demonstrate that Topo-GS provides detailed volumetric visualization without compromising the metric foundation established by the dimensionality-reduction baseline. The visual complexity—such as the trajectory buckling and surface tiling observed in the results—is primarily encoded in the Gaussian covariance tensors, thereby largely preserving the positional distribution of the manifold centers.

\subsection{Ablation Study}
To evaluate the individual contributions of the geometric constraints proposed in our framework, we conducted an ablation study on the Swiss Roll dataset. We isolated the effects of Adaptive Covariance Alignment ($\mathcal{L}_{C}$) and Topological Orientation Smoothing ($\mathcal{L}_{O}$) by disabling them independently. The baseline for comparison is the full Topo-GS pipeline ($\lambda_R=10, \lambda_C=10, \lambda_O=2$), shown in Fig.~\ref{fig:ablation}(a).

\subsubsection{Impact of Covariance Alignment (w/o $\mathcal{L}_{C}$)}
When the covariance alignment constraint is removed ($\lambda_C=0$), the optimization relies solely on the Local Affine Rigidity ($\mathcal{L}_{R}$) anchor. As observed in Fig.~\ref{fig:ablation}(b), the primitives fail to stretch along the local tangent spaces, resulting in an isotropic collapse. The representation degrades into a discrete set of spherical primitives, failing to render the manifold as a cohesive volumetric shell. This empirically confirms that $\mathcal{L}_{C}$ is the primary driver for translating the local spatial footprint into structural geometry.

\subsubsection{Impact of Orientation Smoothing (w/o $\mathcal{L}_{O}$)}
\begin{figure*}[!t]
\centering
\subfloat[Full Topo-GS (Baseline)]{\includegraphics[width=0.32\textwidth]{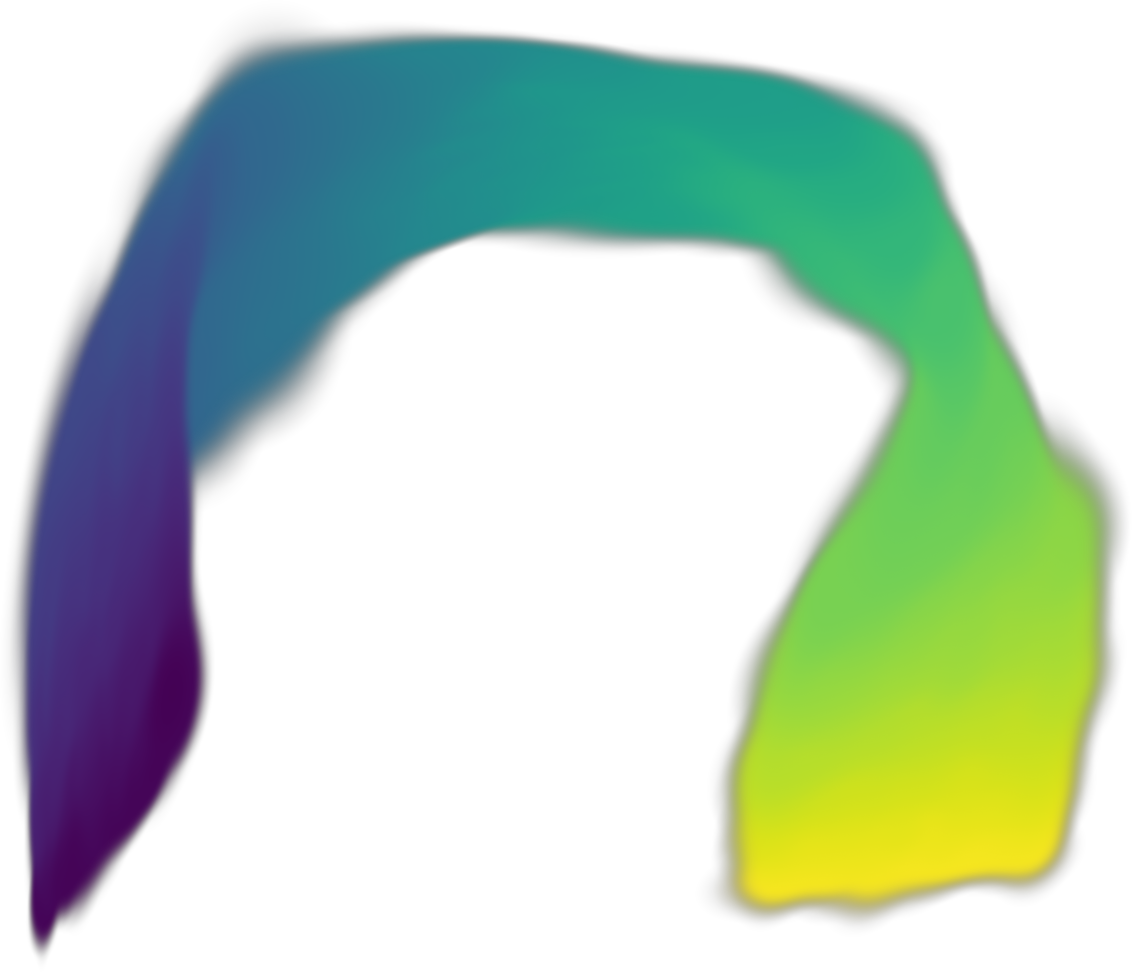}\label{fig:abl_full}}
\hfil
\subfloat[w/o Covariance Alignment ($\mathcal{L}_C=0$)]{\includegraphics[width=0.32\textwidth]{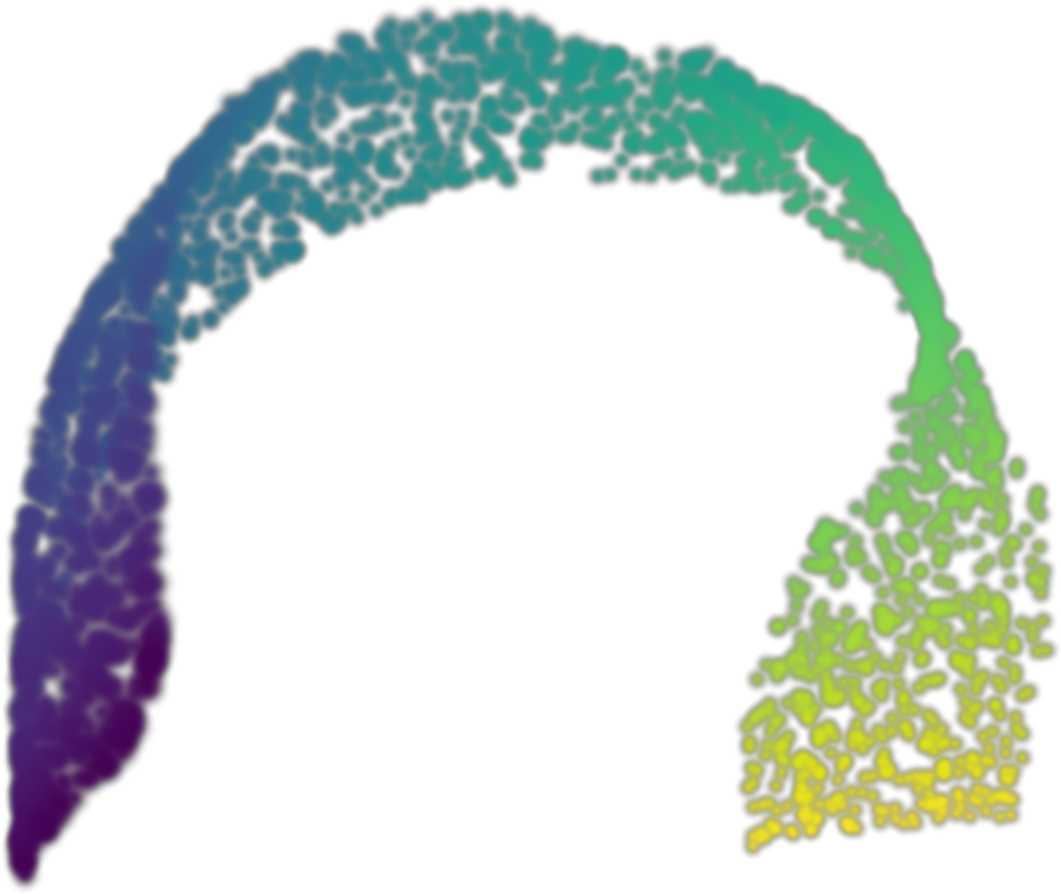}\label{fig:abl_wo_c}}
\hfil
\subfloat[w/o Orientation Smoothing ($\mathcal{L}_O=0$)]{\includegraphics[width=0.32\textwidth]{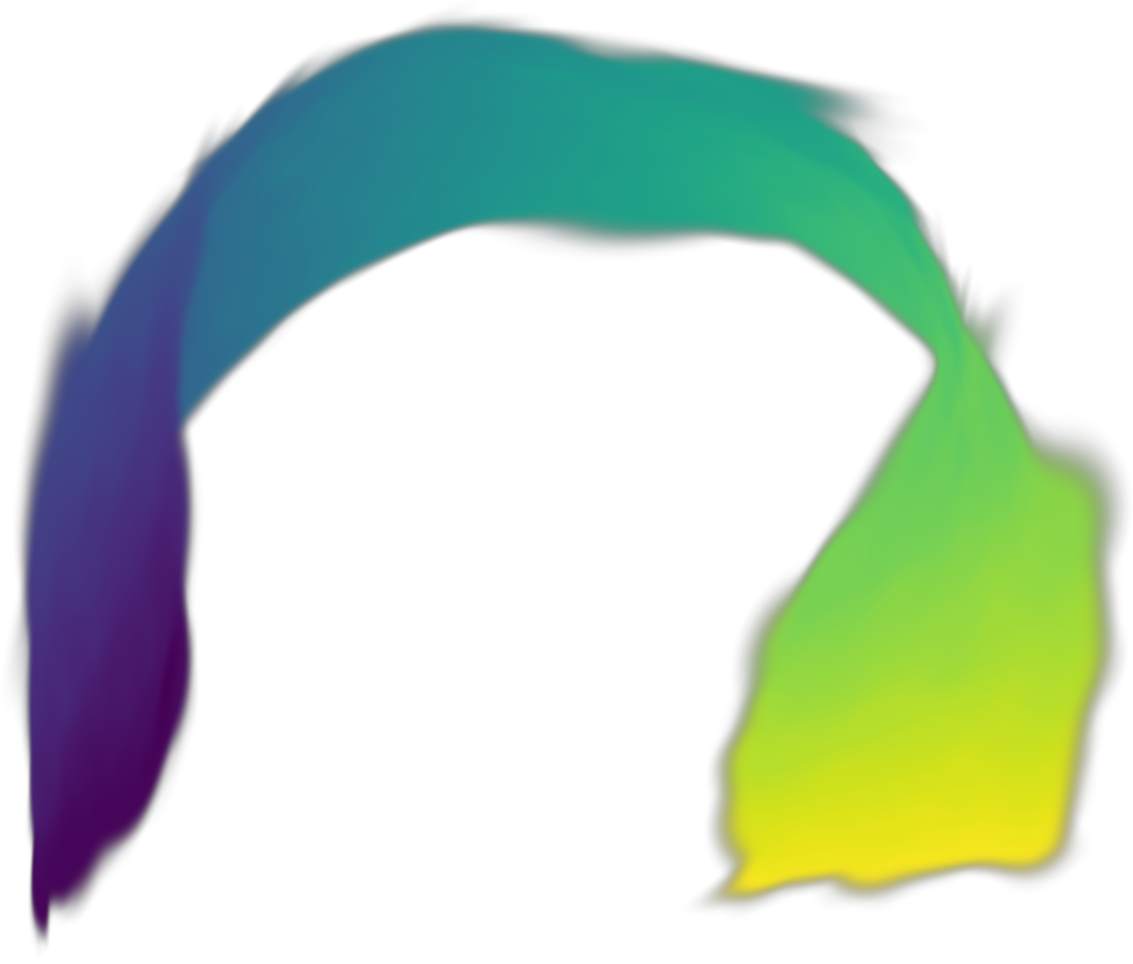}\label{fig:abl_wo_o}}
\caption{Ablation study demonstrating the structural necessity of the proposed geometric constraints. (a) The full Topo-GS engine produces cohesive volumetric tiling. (b) Disabling $\mathcal{L}_C$ results in an isotropic collapse, reducing the representation to a discrete set of spherical primitives. (c) Disabling $\mathcal{L}_O$ results in uncoordinated local rotations, visually manifesting as surface roughness and jagged protruding artifacts along the manifold's boundaries.}
\label{fig:ablation}
\end{figure*}

Disabling the rotational penalty ($\lambda_O=0$) preserves the macroscopic structure but degrades local smoothness. While the primitives successfully flatten into planar shapes driven by $\mathcal{L}_{C}$, the absence of orientation smoothing causes uncoordinated rotations. As shown in Fig.~\ref{fig:ablation}(c), adjacent primitives intersect at misaligned angles, creating visible spiky artifacts and jagged silhouettes along the manifold. This demonstrates that $\mathcal{L}_{O}$ is essential to enforce tangent plane consistency and generate a readable, artifact-free surface.

\section{Discussion and Limitations}
The results demonstrate that Topo-GS effectively bridges the gap between discrete embeddings and continuous volumetric representations. Preserving Trustworthiness and Continuity metrics comparable to the UMAP baseline confirms that introducing geometric constraints does not degrade the underlying projection fidelity. 

However, some limitations remain. The current implementation relies on a $k$-NN graph constructed in the high-dimensional space, which may be sensitive to noise or severe outliers, potentially leading to local artifacts in the Gaussian tiling. Furthermore, while the \emph{lazy SVD} strategy significantly mitigates the computational overhead of the orthogonal Procrustes alignment, the eigendecomposition step remains a theoretical bottleneck for scaling to massive datasets ($N \gg 10^5$). Future work will explore hierarchical optimization schemes and the integration of density-based opacity mapping to further enhance the visual resolvability of multi-scale manifold structures.

\section{Conclusion}
We presented Topo-GS, a dimensionality reduction framework that repurposes 3D Gaussian Splatting as a geometric projection engine. By transitioning from the traditional discrete point-cloud paradigm to continuous volumetric primitives, Topo-GS fundamentally mitigates visual occlusion and spatial fragmentation in high-dimensional data analysis.

Central to our approach is the integration of topology-aware constraints. We demonstrated that effective manifold unrolling requires explicit geometric priors, notably the alignment of spatial covariances and the application of dataset-specific rigidities—such as robust stress absorption for 1D trajectories and strict Procrustes targets for cohesive 2D surfaces. Quantitative evaluations and ablation studies confirm that Topo-GS successfully renders volumetric continuity while preserving the local topological fidelity of baseline algorithms such as UMAP. Consequently, inherent projection distortions are no longer obscured but explicitly manifested as observable variations in geometric thickness and structural buckling.

Future work will extend this framework to time-varying manifolds, investigate hierarchical optimization schemes to accelerate Procrustes alignment for massive datasets, and integrate these continuous representations into immersive spatial visualization platforms.

\section*{Acknowledgments}
This study was financed in part by the Coordenação de Aperfeiçoamento de Pessoal de Nível Superior - Brasil (CAPES) - Finance Code 001, and Conselho Nacional de Desenvolvimento Científico e Tecnológico (CNPq) (403280/2025-7).

\section*{AI Assistance Declaration}
In compliance with scientific integrity guidelines, the authors declare the use of Grammarly Pro and Google Gemini during the preparation of this manuscript. These AI tools were employed exclusively to assist with language refinement, LaTeX formatting, and code refactoring. The authors reviewed, edited, and validated all AI-generated text and code, and assume full responsibility for the accuracy, originality, and final content of this publication.

\bibliographystyle{IEEEtran} 
\bibliography{ref}
\end{document}